# Finite-Size analysis of the 4-d abelian surface gauge model

M. Baig[1] and R. Villanova[1,2]

[1] Grup de Física Teòrica
Institut de Física d'Altes Energies
Universitat Autònoma de Barcelona
08193 Bellaterra (Barcelona), Spain

[2] Matemàtiques Aplicades
Escola de Ciències Empresarials
Universitat Pompeu Fabra
Rambla de Santa Mònica, 32
08002 Barcelona, Spain

### Abstract

We present the results of a finite-size analysis of the four dimensional abelian surface gauge model. This model is defined assigning abelian variables to the plaquettes of an hypercubical lattice, and is dual to the four dimensional Ising model. This last model is known to present a second order phase transition with mean field critical exponents. We have performed Monte Carlo simulations on several lattice sizes and high statistics. The analysis of the partition function zeroes and the specific heat scaling behaviour allowed us to estimate the critical coupling $\beta_c$ as well as the critical exponents $\nu$ and $\alpha$. Our results are consistent with the second order critical exponents $\nu = 1/2$ and $\alpha = 0$. The $\beta_c$ value is in perfect agreement with duality predictions from the 4-d Ising model. Nevertheless, the energy histograms show a seemingly non-vanishing double peak structure. The interface tension analysis suggests that this may be a finite size effect.

---

[2] Current address.

**Introduction**

The abelian surface gauge model was proposed some time ago in order to have a natural lattice translation of the three-index fields that appear in theories based on an antisymmetrical potential [1] $\phi(x)_{\mu\nu}$. Such theories are relevant in different domains as Chern-Simons studies, supergravity or the 10-dimensional $E_8$ gauge theories [2]. Although different lattice strategies have been proposed [3], the surface gauge model seems to be a simple and numerically attainable way to study these theories.

In a recent paper [4] a numerical analysis of the surface gauge model for the abelian Z(2) gauge group was performed for different space-time dimensionalities. In summary, the model is exactly solvable in three dimensions showing no phase transitions, evidences for a second order phase transition were found for the four dimensional case, whereas it was realized the existence of a first order phase transition for higher dimensionalities, d=5 and 6. Moreover, duality relations with several spin models were also studied. In particular, the four dimensional case is dual to the four dimensional Ising spin model, which is known to have a second order phase transition with mean field critical exponents.

In this paper we present a high statistics finite-size analysis of the phase transition for d=4 on larger lattices. Our main purpose is to confirm the previous work, performed on rather small lattices, and to estimate the critical exponents of the theory.

**The model**

We denote by $n$ a lattice site and by $n_\mu$ a link starting from point $n$ in the direction $\mu$ of an 4-d hypercubical lattice. Starting from this point there are 4 positive links. A simple plaquette defined by directions $\mu, \nu$ is denoted by $n_{\mu\nu}$. We assign a gauge variable ($\sigma_i = \pm 1$) to each plaquette. Elementary three-dimensional cubes are composed by six plaquettes ($i \in \partial_{cube}$, i.e. the plaquette belongs to the perimeter of the cube). Three of them are the plaquettes associated to the combinations of the three index $\mu, \nu, \eta$ related to the links starting from a given point $n$ in the positive directions. The remainder three plaquettes are those that "close" the cube. The partition function is defined, then, as:

$$Z = \sum_{\{\sigma_i\}} e^{-\beta E}; \qquad E = \sum_{cube=1}^{DL^D} E(cube); \qquad (1)$$

where

$$E(cube) = 1 - \prod_{i \in \partial_{cube}} \sigma_i; \qquad \sigma_i = \pm 1, \qquad (2)$$

with $\beta > 0$ being the inverse temperature in natural units. From now on, $< . >$ will denote thermal averages of an observable in the canonical distribution.

**Duality**

A duality transformation [5] relates the four dimensional surface gauge model to the four dimensional Ising model. This last model has a second order phase transition with mean field critical exponents [6]. From series analysis [7], the critical temperature of the four dimensional Ising model has been estimated to be $\beta_c^* = 0.149\,65 \pm 0.000\,05$. This value is in agreement with the numerical simulations of [6].

The duality transformation between these models implies the following relation between the critical couplings

$$\beta_c = \frac{1}{2} \ln \frac{e^{2\beta_c^*} + 1}{e^{2\beta_c^*} - 1}, \qquad (3)$$

where $\beta_c^*$ is the critical coupling of the four dimensional Ising model and $\beta_c$ is the surface gauge model one.

From this equation, one can estimate the numerical value of the critical temperature for the surface gauge model giving $\beta_c = 0.953\,44 \pm 0.000\,16$, value in accordance with the numerical simulations of [4].

The finite size analysis of [6] found, for the 4-d Ising model, a second order phase transition with critical exponents $\alpha = 0.04 \pm 0.06$ and $\nu = 0.510 \pm 0.016$, i.e. compatible with zero and with $1/2$ respectively. These values for the critical exponents are, then, compatible with those expected from mean field. In his turn, duality implies also mean field critical exponents for the surface gauge case. Finally, it should be noticed that the Ising model has a local order parameter, the magnetization, that, due to the gauge invariance, is not present in the gauge version. This reduces the number of critical exponents to be measured.

## Monte Carlo calculation

Our numerical simulations with the surface gauge model have been performed applying the Metropolis algorithm [8] as it was done in Ref [4]. The Monte Carlo simulation took place for the following lattices and MC $\beta$'s:

$$
\begin{aligned}
(L = 3; \quad \beta &= 0.850), \\
(L = 5; \quad \beta &= 0.920), \\
(L = 6; \quad \beta &= 0.920,\ 0.930), \\
(L = 9; \quad \beta &= 0.940,\ 0.943), \\
(L = 12; \quad \beta &= 0.9474).
\end{aligned}
\tag{4}
$$

To reach equilibrium, the first 50 000 sweeps were discarded from an initially ordered state, and further (from 500 000 to 1 000 000) sweeps were generated for measurements. After each sweep, the energy was measured and recorded in a time series file. Fig. 1a shows the first 50 000 energies recorded for $L = 9$ at $\beta = 0.943$. We checked that our results didn't depend on the initial configuration, studying how they changed when we discarded some of the bins into which we splitted the data. Besides, for $L = 5$ we repeated the simulation with a hot start.

From the time series of $E$, it is straightforward to compute in the FSS region various quantities at nearby values of the Monte Carlo (MC) $\beta$'s by standard Ferrenberg-Swendsen reweighting [9]. To estimate the statistical errors, the time-series data was split into 20 bins, which were jack-knived [14] to decrease the bias in the analysis of the reweighted data.

From the previous work in [4], and some new test runs, the above MC $\beta$'s were known to be close to the (lattice dependent) effective critical couplings, $\beta_c(L)$, of the system. We understand by effective critical beta the one at which the specific heat of a finite system has a peak.

In the two cases ($L = 6, 9$) in which our first MC run was not close enough to the effective critical $\beta_c(L)$ of the system, we used the Ferrenberg-Swendsen reweighting procedure to determine it more precisely, and then launched a second MC run at this newly found effective critical beta.

In order to combine the information contained in the two MC runs (at different $\beta$'s) that took place for $L = 6$ and the two for $L = 9$, we used a patching procedure introduced in [10]. This patching procedure allows to estimate a single density of states from several MC runs by means of reweighting and least-square fitting of the energy histograms of the runs. Once the procedure has been applied, one can investigate the critical properties of the model in the vicinity of the phase transition.

Fig. 1b shows the energy histograms of our Monte Carlo simulations ($L = 9$; $\beta = 0.94$) and ($L = 9$; $\beta = 0.943$). Fig. 1c shows the probability distribution $P_L(E;\beta)$ at $\beta = 0.9432$ that we obtain from the energy histograms in Fig. 1b by means of the patching procedure and the reweighting techniques already mentioned.

**Partition function zeroes**

From equations (1)-(2), we see that the energy of our model changes in steps of 4 within the interval $0 \leq E \leq 8\,L^4$. Defining the new variable

$$u = \exp(-4\beta), \tag{5}$$

the partition function becomes, then, a polynomial of degree $2L^4 + 1$ in u. Using the Newton-Raphson method, as described in [11], we calculated the two closest zeroes to the real axis of the complex u-plane. The zeroes are given in Table 1. We have denoted the zeroes by $u_i^0$, where the subindex $i$ stands for the different zeroes, ordered according its distance to the real axis. For instance, $u_1^0$ corresponds to the zero with the smallest imaginary part.

Let us define $u_c = u(\beta_c)$. It was first shown in [12] that, for $L$ big enough, the finite-size scaling analysis of the partition function zeroes is given by

$$u_i^0(L) = u_c + AL^{-1/\nu}[1 + O(L^{-\omega})], \quad \omega > 0, \tag{6}$$

where A is a complex constant.

This result has been used both for first and second order phase transitions as well as spin systems and gauge theories [13, 10].

For the present case we use the imaginary part of the zeroes only. It turns out to be more effective. The reason is that the $L$ dependence of the real part is rather weak and consequently has a worse signal-to-noise ratio than the imaginary part. This makes (6) less favorable, in this case, than just its imaginary part.

Table 1: *Table of the $1^{st}$ and $2^{nd}$ zeroes of the partition function*

| $L$ | $Re(u_1^0)$ | $Im(u_1^0)$ | $Re(u_2^0)$ | $Im(u_2^0)$ |
|---|---|---|---|---|
| 3 | 0.031630(34) | 0.008984(14) | 0.03143(12) | 0.020219(92) |
| 5 | 0.025197(18) | 0.0022220(97) | 0.026316(72) | 0.004851(61) |
| 6 | 0.024098(21) | 0.0014268(86) | 0.024905(43) | 0.003152(37) |
| 9 | 0.022963(11) | 0.0005793(56) | 0.024205(68) | 0.001139(12) |
| 12 | 0.022566(12) | 0.0003133(65) | 0.022789(42) | 0.000643(32) |

Due to the duality relations between our model and the 4d Ising model, we expect to obtain a result compatible with $\nu = 0.5$.

In first approximation one can resort to

$$Im(u_1^0(L)) = Im(A)L^{-1/\nu}, \qquad (7)$$

which for pairs of lattices allows to estimate $\nu(L, L')$ from

$$\nu(L, L') = \frac{\ln(L'/L)}{\ln\left(Im(u_1^0(L'))/Im(u_1^0(L))\right)} \ . \qquad (8)$$

Our results are shown in Table 2. For increasing $\min(L, L')$ there seems to be a trend $\nu(L, L') \to 0.5$, as expected from duality. Our best estimate is (matching our largest lattices $L = 9$ and $12$)

$$\nu = 0.469(17), \qquad (9)$$

which is consistent with 0.5 within two error bars.

On the other hand, a linear regression fit of (7) with $L = 6, 9$, and $12$ gives us

$$\nu = 0.4531(46) \qquad (10)$$

with a goodness of the fit [11] of $Q = 0.36$. The lattices $L = 3$ and $5$ worsen the fit.

Table 2: *Estimates of $\nu(L, L')$ for pairs of lattices of size $L$ and $L'$*

| $L \setminus L'$ | 5 | 6 | 9 | 12 |
|---|---|---|---|---|
| 3 | 0.3657(12) | 0.3768(13) | 0.4008(14) | 0.4131(26) |
| 5 |  | 0.4116(69) | 0.4372(34) | 0.4469(48) |
| 6 |  |  | 0.4498(57) | 0.4572(65) |
| 9 |  |  |  | 0.469(17) |

A first order phase transition would have a critical exponent $\nu = 1/d = 0.25$ ($d$ : spatial dimension). Our estimation of $\nu$, then, excludes the possibility of a first order transition.

**Specific heat**

As was mentioned before, we compute quantities at nearby values of the MC $\beta$'s by standard reweighting, and use jack-knived bins to take care of the statistical errors. In this fashion we determine the maxima of the specific heat

$$C(\beta) = \frac{\beta^2}{N_c}(<E^2> - <E>^2), \qquad (11)$$

where $N_c$ is the number of three-dimensional cubes in lattice.

The results are shown in Table 3, together with their effective critical $\beta$'s.

Let us also define $\beta^{zero}(L)$ for the first zero of the partition function. From eq. (5) we have

$$\beta^{zero}(L) = -\frac{1}{8} \ln\left[Re(u_1^0(L))^2 + Im(u_1^0(L))^2\right]. \qquad (12)$$

It can be considered as another effective critical $\beta_c(L)$. Table 3 compares it, for different lattice sizes, with the specific heat effective critical $\beta$'s.

For a second order phase transition, the locations of the specific heat (or any other set of effective critical $\beta$'s) are expected to scale as [15]

$$\beta_{max}^C(L) = \beta_c + aL^{-1/\nu}, \qquad (13)$$

Table 3: *Table of the effective critical $\beta^{zero}$, the specific heat ($C_{\max}$), and its pseudo critical coupling*

| L | $\beta^{zero}$ | $\beta^C_{\max}$ | $C_{\max}$ |
|---|---|---|---|
| 3 | 0.85371(25) | 0.85277(25) | 1.7040(41) |
| 5 | 0.91929(18) | 0.91825(17) | 2.514(18) |
| 6 | 0.93097(22) | 0.92907(20) | 2.772(23) |
| 9 | 0.94339(12) | 0.94300(11) | 3.106(40) |
| 12 | 0.94780(13) | 0.94760(13) | 3.277(97) |

with $a$ being a constant. If the temperature-driven transition of the model happened to be of first order, the scaling would have the same form, with $\nu$ replaced by $1/d$ ($d$: spatial dimension).

Assuming $\nu = 1/2$, we can use eq. (13) to obtain estimates of the critical coupling $\beta_c(\infty)$ from linear fits in $1/L^2$. Our results from fits to the data (see Table 3) of the four and three largest lattices respectively are

$$\beta_c = 0.95390 \pm 0.00012 \quad \text{(from } \beta^C_{max}\text{)} \tag{14}$$
$$\beta_c = 0.95389 \pm 0.00016 \quad \text{(from } \beta^{zero}\text{)}. \tag{15}$$

If instead of assuming $\nu = 1/2$ we use our best estimate $\nu = 0.469$ from eq. (9), the fits yield

$$\beta_c = 0.95294 \pm 0.00012 \quad \text{(from } \beta^C_{max}\text{)} \tag{16}$$
$$\beta_c = 0.95268 \pm 0.00016 \quad \text{(from } \beta^{zero}\text{)}. \tag{17}$$

The agreement between these results and the estimate based on the duality relation with the 4d Ising model, $\beta = 0.95344(17)$, is slightly better when taking $\nu = 1/2$.

We now turn to the specific heat which is usually the most difficult quantity to analyze. The reason is that the critical divergence is generally rather weak and regular background terms become important. Recalling our $\nu \approx 0.5$ and assuming hyperscaling to be valid for $d = 4$ as well, we expect

$$\alpha = 2 - d\nu \approx 0. \qquad (18)$$

The corresponding FSS prediction is then [15]

$$C_{max}(L) = C(\beta^C_{max}(L), L) = B_0 + B_1 \ln L. \qquad (19)$$

A linear fit with the three largest lattices gives $B_0 = 1.36(18)$ and $B_1 = 0.790(94)$ with $Q = 0.6$. See Fig. 2.

In should be remarked, however, that the confirmation of $\alpha = 0(\log)$ is not really conclusive. Due to the small range over which $C_{max}$ varies we can fit the data also with a simple power-law Ansatz $C_{max} \propto L^{\alpha/\nu}$, yielding $\alpha/\nu = 0.265(31)$ with $Q = 0.46$ [from eq. (18) we have $\alpha/\nu = 2/\nu - d$, and using our best estimate $\nu = 0.469(17)$, results in $\alpha/\nu = 0.26(16)$]. We also tried a non-linear three-parameter fit to the more reasonable Ansatz $C_{max} = b_0 + b_1 L^{\alpha/\nu}$, but our data was not good, and the error bars turn out to be too large to draw any conclusion.

Notice that in a first order phase transition we would expect $\alpha/\nu = d = 4$.

**Interfacial free energy**

The Ferrenberg-Swendsen reweighting procedure [9] allows to estimate the canonical probability distribution $P_L(E; \beta)$ for a given lattice size and $\beta$ from the energy histogram $hist(E; \beta_{MC})$ obtained in the Monte Carlo simulation (if our $\beta$ of interest is close enough to the simulation $\beta$). First order phase transitions are characterized by a double peak structure in the probability density $P_L(E; \beta)$ sufficiently close to the transition point.

Fig. 1c shows our estimated probability distribution for $L = 9$ at $\beta = 0.9432$. It has a typical first order transition double peak estructure. All our lattice sizes suffer the same effect. At the same time, we know from the previous sections that our estimates for the critical exponents $\nu$ and $\alpha$ exclude a first order phase transition, and are consistent with a 4d Ising model like second order phase transition.

In order to analyze this phenomena, we adopt the normalization $P_L^{max} = 1$, and define an effective critical point $\beta_c^P$ by the requirement that both maxima are of equal height. If $E_L^{1,max} < E_L^{2,max}$, we have, then:

$$P_L^{1,max} = P_L(E_L^{1,max}; \beta_c^P) = P_L^{2,max} = P_L(E_L^{2,max}; \beta_c^P) = 1. \qquad (20)$$

We introduce also $P_L^{min} = P_L(E_L^{min}; \beta_C^P)$ as the minimum probability density between the two peaks. Fig. 1c shows it for $L = 9$.

Following [16], we argue that in order to be in the presence of a first order phase transition, $P_L^{min}$ has to decrease as $L$ grows. More specifically, for a first order phase transition the interfacial free energy per unit area (or surface tension $f^s$) separating the coexisting ordered and disordered phases is given [17] by the limit $L \to \infty$ of the quantity

$$2f_L^s = -\frac{1}{L^{d-1}} \ln(P_L^{min}). \tag{21}$$

Table 4 presents our results for $P_L^{min}$ and $f_L^s$ together with their corresponding $\beta_c^P$.

Table 4: *Table of $\beta_c^P$, $P_L^{min}$ and the corresponding $2f_L^s$*

| $L$ | $\beta_c^P$ | $P_L^{min}$ | $2f_L^s$ |
|---|---|---|---|
| 3 | 0.83870(22) | 0.4208(52) | 0.03206(46) |
| 5 | 0.91885(22) | 0.609(32) | 0.00397(42) |
| 6 | 0.92939(30) | 0.609(37) | 0.00230(28) |
| 9 | 0.94320(10) | 0.557(24) | 0.000804(60) |
| 12 | 0.94773(17) | 0.543(93) | 0.000355(69) |

Following [17, 18] we perform the FSS fit

$$2f^s = 2f_L^s + c/L^{d-1} \tag{22}$$

with our four larger lattices in order to extract the $L \to \infty$ limit of eq. 21. It is depicted in Fig. 3 and gives

$$2f^s = 0.000103 \pm 0.000070, \tag{23}$$

with $c = -0.491 \pm 0.045$ for the constant, and a goodness of the fit $Q = 0.8$. Notice that this result is compatible with $2f^s = 0$ within 2 error bars. In fact, we have also tried the fit

$$2f_L^s = -c/L^{d-1}, \tag{24}$$

and obtained $c = -0.541 \pm 0.029$ with $Q = 0.4$.

Based on these results, we argue that, in the infinite volume limit, the surface tension disappears, and we are left with a continuous phase transition in the critical point $\beta_c$, as we expected from duality.

**Conclusions**

We have performed a fairly detailed Monte Carlo analysis of the 4-d abelian surface gauge model. Our estimates for the critical coupling ($\beta_c = 0.95390(12)$, from the peak location of the specific heat, and $\beta = 0.95389(16)$ from the closest zero to the real axis) agree with the estimate based on the duality relation with the 4d Ising model.

From the finite-size scaling analysis of the first partition function zeroes we have extracted critical exponents $\nu$ which discard a first order phase transition, and that manisfestly show a trend towards the $\nu = 0.5$ 4d Ising result. Our best estimate is $\nu(9, 12) = 0.469(17)$.

The analysis of the specific heat peaks is more problematic because our data is compatible with a critical exponent $\alpha = 0$ (as should be expected from duality), but also with a power-law scaling $\alpha \neq 0$. In fact, we are unable to discriminate between the two Ansatz unless we go to larger lattices ($L = 16$ would probably be enough).

Altogether, considering our estimates for $\beta_c$, $\nu$, and $\alpha$, we can exclude a first order phase transition. It is arguable that they are a good Monte Carlo evidence in support of the duality relation between the surface gauge model and the 4d Ising model.

The typical double peak structure that our probability densities present are somehow puzzling, because as we have seen a 4d Ising model like second order phase transition seems to hold for the model. A further finite-size analysis of the $P_L^{min}$'s reveals that in the limit $L \to \infty$ the surface tension is compatible with zero, giving place to a continuous transition.

We conclude, then, that the system experiments actually a second order phase transition with mean field critical exponents, as predicted by duality. The intriguing behaviour that originates the double peak structure seems to be due to an added phenomena that may be understood in terms of the topological excitations of the model, i.e. the monopoles. In the QED case, these structures are 1-dimensional, and it has recently been noticed [19] that the imposition of periodic boundary conditions originate an unphysical stability of these loops on the hypertorus, inducing a first order transition. In the present case, however, the topological excitations are point-like and this effect will not be present. Nevertheless, these excitations can manifest other phenomena, as condensation or site percolation, that can be responsible of the strange metastability over the phase transition. Note that this image is consistent with the fact than in five dimensions, where the topological excitations will become 1-dimensional, this model exhibits a clear first order phase transition. Work is in progress on this direction.

**Acknowledgement**

Numerical simulations have been performed mainly on the cluster of IBM RISC-6000 workstations of IFAE and on the CRAY-YMP of CESCA. The support of CESCA and CICYT (contract number AEN 93-0474) are acknowledged. M.B. would like to thank Hugo Fort and R.V. Wolfhard Janke for helpful discussions.

**Figure Captions**

Fig. 1a Time history of the first 50 000 energy measurements at $\beta = 0.943$ on the $L = 9$ lattice.

Fig. 1b Energy histograms of our simulations $(L = 9;\ \beta = 0.94)$ and $(L = 9;\ \beta = 0.943)$.

Fig. 1c Canonical probability distribution $P_L(E;\beta)$, normalized by $P_L^{1,max} = P_L^{2,max}$ at $\beta = 0.9432$. It has been estimated from the energy histograms in Fig. 1b by means of the patching procedure introduced in [10] and the reweighting techniques of [9].

Fig. 2 Finite-size scaling analysis of the specific-heat maxima $C_{max}$. Also shown are least-squares fits to a logarithmic Ansatz, $C_{max} = B_0 + B_1 \ln L$ (with $B_0 = 1.36(18)$, $B_1 = 0.790(94)$), and to a pure power-law Ansatz, $C_{max} = cL^{\alpha/\nu}$ (with $c = 1.73(10)$, $\alpha/\nu = 0.265(31)$).

Fig. 3 Finite-size scaling fit2, $2f^s = 2f_L^s + c/L^{d-1}$ (with $c = -0.491(45)$), and $0 = 2f_L^s + c/L^{d-1}$ (with $c = -0.541(29)$), from the four larger lattices for the interfacial free energy. The dashed straight line is the asymptotic limit $2f^s = 0.000103$ obtained in the first case.

**Fig. 1 can be requested via E-mail to baig@ifae.es or villanova@ifae.es**

Fig. 1

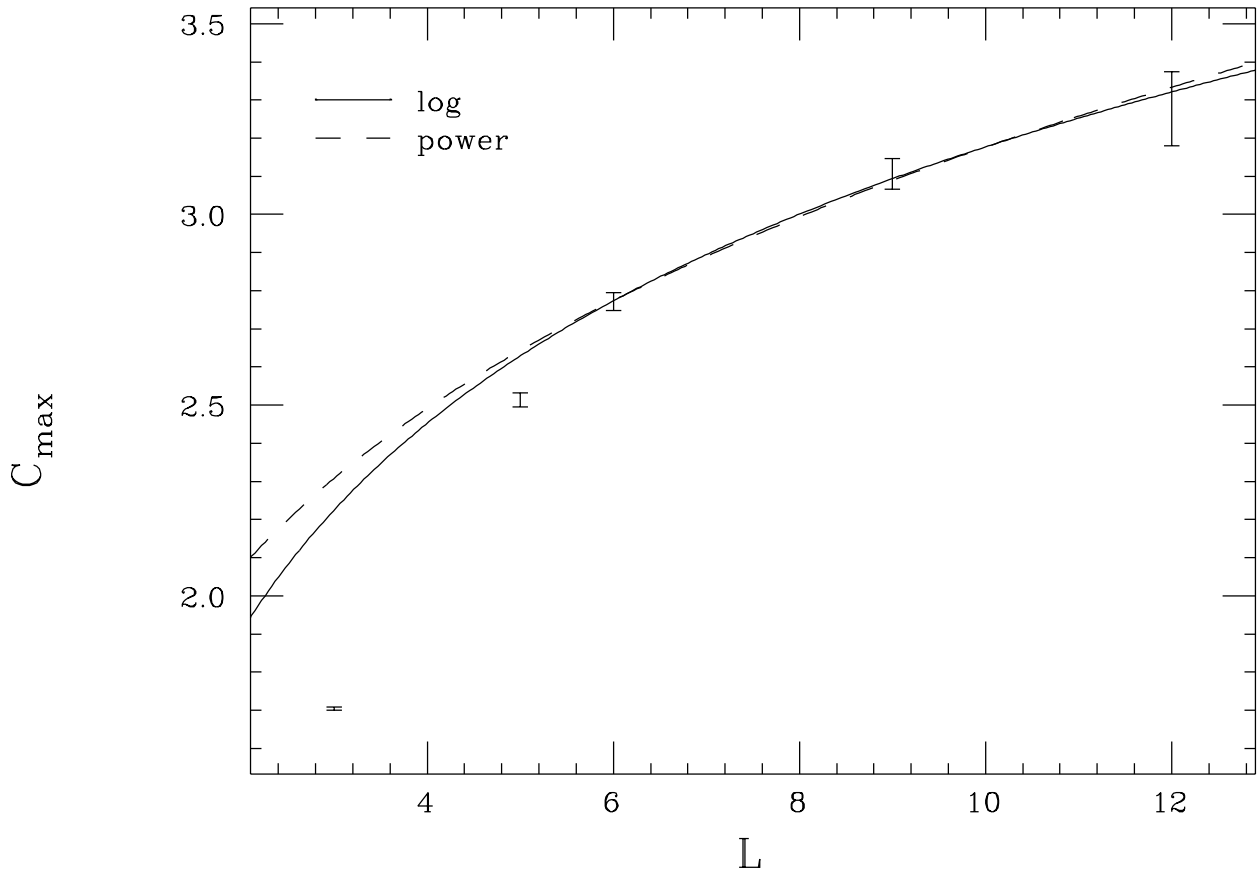

Fig. 2

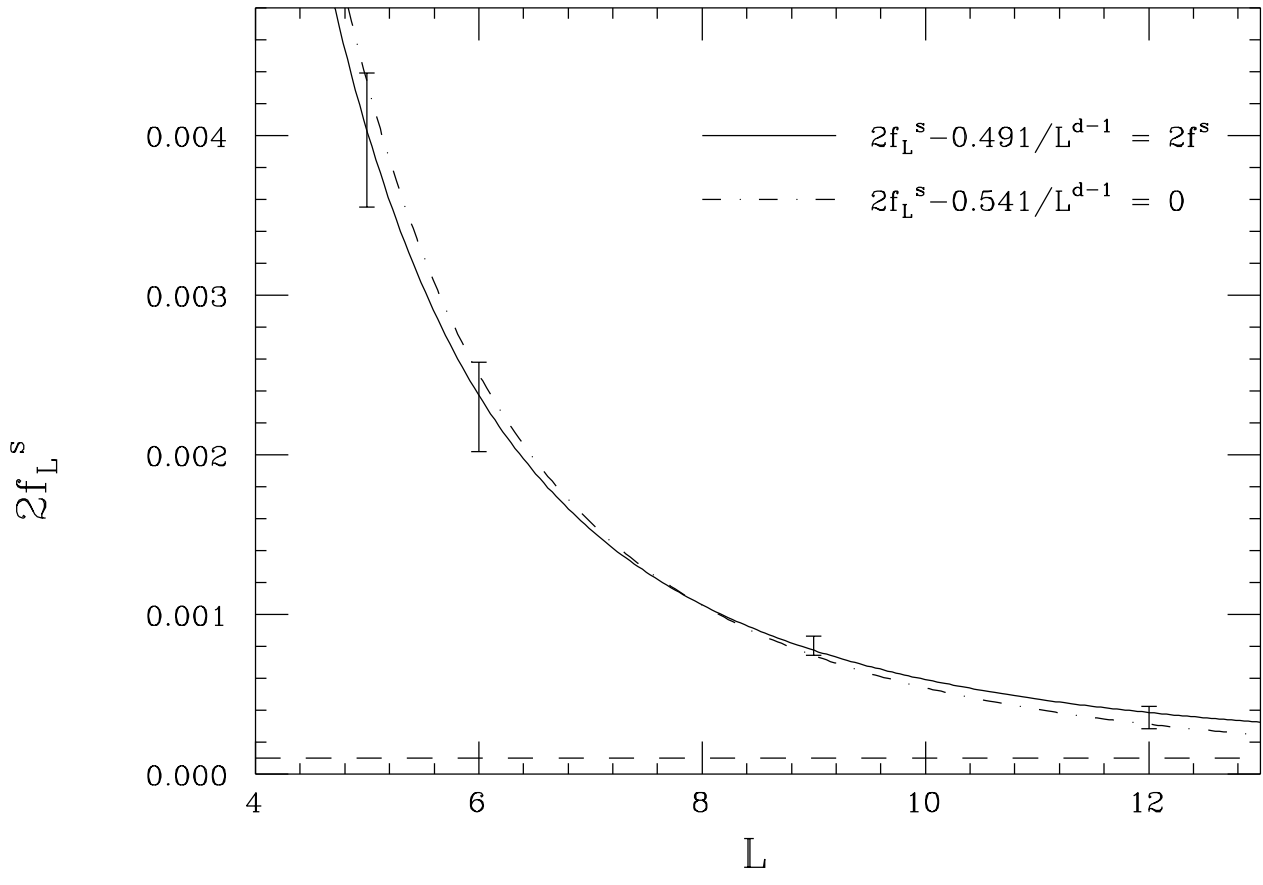

Fig. 3